\documentclass{article}
\usepackage{spconf,amsmath,graphicx}
\usepackage{booktabs}
\usepackage{subcaption}
\usepackage{enumitem,kantlipsum}
\title{Beyond Empirical Windowing: An Attention-Based Approach for Trust Prediction in Autonomous Vehicles}
\name{Minxue Niu$^{2}$\sthanks{work completed during internship at HRI}, Zhaobo K. Zheng$^1$, Kumar Akash$^1$, Teruhisa Misu$^1$\vspace{-4pt}}
\address{$^1$Honda Research Institute USA, Inc., San Jose, California, USA\\ $^2$ University of Michigan, Ann Arbor, Michigan, USA\vspace{-15pt}}

\begin{document}
\ninept
\maketitle

\begin{abstract}
%Recent advancements in multimodal sensing and machine learning techniques have enhanced human state detection, which 
Humans' internal states play a key role in human-machine interaction, leading to the rise of human state estimation as a prominent field. Compared to swift state changes such as surprise and irritation, modeling gradual states like trust and satisfaction are further challenged by label sparsity: long time-series signals are usually associated with a single label, making it difficult to identify the critical span of state shifts. Windowing has been one widely-used technique to enable localized analysis of long time-series data. However, the performance of downstream models can be sensitive to the window size, and determining the optimal window size demands domain expertise and extensive search. To address this challenge, we propose a Selective Windowing Attention Network (SWAN), which employs window prompts and masked attention transformation to enable the selection of attended intervals with flexible lengths. We evaluate SWAN on the task of trust prediction on a new multimodal driving simulation dataset. Experiments show that SWAN significantly outperforms an existing empirical window selection baseline and neural network baselines including CNN-LSTM and Transformer. Furthermore, it shows robustness across a wide span of windowing ranges, compared to the traditional windowing approach.
\end{abstract}
\vspace{-5pt}
\setlength{\abovedisplayskip}{5pt}
\setlength{\belowdisplayskip}{5pt}
\section{Introduction}
\label{sec:intro}\vspace{-5pt}
Users' emotional and cognitive conditions deeply influence how they interact with machine systems and are closely connected to safety, system efficiency, and user experience. With the advances in multimodal sensing and machine learning, research on human state modeling has become an increasingly prominent area~\cite{zhang2020emotion}. 
% Many work have focused on emotion analysis in Human-Machine Interaction (HMI)~\cite{cowie2001emotion, abdullah2021multimodal}, where multimodal time-series signals are segmented into short time spans usually ranging from xx to xx seconds~\cite{}.
%Emotion analysis has been extensively studied, with application in  chatbots~\cite{pamungkas2019emotionally}, driving assistance~\cite{zepf2020driver}, healthcare and smart home systems~\cite{mano2016exploiting}, .etc. 
Unlike short-term states such as emotions~\cite{cowie2001emotion, abdullah2021multimodal}, modeling protracted human states like satisfaction and trust presents unique challenges due to long signals and sparse labeling. For example, research on trust in Automated Vehicles (AVs) reveals that user trust is generally stable, with variations only at specific critical events~\cite{ekman2017creating}. Frequent surveys to capture these variations can be a major distraction for participants~\cite{porter2004multiple}. This results in long input data sequences characterized by infrequent spans that signify state changes with output being a single label. In this work, we tackle this challenge in the application of user trust modeling in AVs. Trust in automation affects user anxiety~\cite{kraus2020scared}, preferred driving style ~\cite{zheng2022identification}, wellbeing \cite{li2023you}, as well as the acceptance and use of AV systems~\cite{molnar2018understanding}. It's challenging yet important to identify localized signal patterns of trust change over a long time-series.

Windowing has been one of the most common signal processing techniques for time-series data~\cite{oppenheim1999discrete}. It enables localized analysis through the segmentation of a prolonged sequence into fixed-size, often overlapping intervals. In the literature of AV trust modeling, it's a common approach to extract features from multimodal sensing signals with a fixed-size sliding window, and perform classification with traditional machine learning classifiers~\cite{akash2018classification, ayoub2022real, konishi2022inferring}.
%For example, Akash et al. used physiological signals to sense trust, combining windowing with feature extraction and a quadratic discriminant classifier~\cite{akash2018classification}. Similarly, Ayoub et al. applied windowing with an XGBoost classifier on eye gaze and heart rate signals~\cite{ayoub2022real}. Konish et al. combined windowing with Random Forest classifier~\cite{konishi2022inferring}.
Despite promising results, such systems' performance is often sensitive to the choice of window sizes. The optimal window size can vary a lot across datasets and models: for example, Akash et al. used a time window of 1 second, while Ayoub et al. used 25 seconds~\cite{akash2018classification, ayoub2022real}. As a result, it takes domain expertise and extensive search to find the optimal window in each dataset. 

Alternatively, Deep Neural Networks (DNNs) such as Convolutional Neural Networks (CNNs) and Transformers have gained traction in multimodal signal processing~\cite{kiranyaz20191, wen2022transformers}. Yi et al. applied a CNN-LSTM network to multimodal trust modeling~\cite{yi2023measurement}. DNNs show remarkable representation and data-mining potential, but their capability of modeling long time-series data is limited~\cite{zeng2023transformers}. For example, transformers calculate pair-wise attention across all steps in each layer, which is not efficient for learning localized features. This issue can be critical with relatively small datasets, which is often the case in the field of human state modeling~\cite{cambria2017affective, liao2022driver}. Transformers with local attention windows have been adopted in computer vision to enable more efficient feature learning~\cite{liu2021swin}, while recent work in speech processing applied attention windows to align modalities~\cite{zhao23b_interspeech}. Their windowing approaches are either limited to self-attention, which doesn't address the challenge of aggregating information from long sequences, or applied in cross-attention between two modalities, which is hard to scale to multiple modalities.
% Some ongoing work utilize LSTM with sliding attention for this challenge, but they focus more on feature fusion.
% A few work has explored attention with limited range or windowed attention. However, 

In this work, we introduce the Selective Windowing Attention Network (SWAN), which combines the intuitive efficacy of windowing and the flexibility of the attention mechanism. With a limited-range localized attention, it mimics the sliding window approach while enabling flexible attention and reducing the sensitivity to window size. 
% Leveraging the attention mechanism, it enables flexible-length attended area selection and reduces the sensitivity to window size. By applying windowing-based masking, it encourages localized feature recognition and greatly enhances training efficiency. 
We contribute a new public dataset focused on user trust in AVs, which includes multimodal contextual and user behavioral signals. On the task of trust prediction, SWAN significantly outperforms empirical windowing and neural network baselines. Through both quantitative and qualitative evaluation, we show that its attention assignment aligns with human expectation, and is very robust across a wide span of window sizes.

\vspace{-6pt}
\section{Dataset}\label{sec:data}\vspace{-5pt}
Existing works on trust prediction in AVs have adopted different datasets, modalities, stimuli, and evaluation methods, making it hard to compare the modeling methods. To address the lack of public datasets in this domain, we collected a new dataset for modeling user trust in AVs, which is available with more detail on the Honda Research Institute website for research purposes. 
% We will make this dataset publicly available upon request for research purposes, upon the publish of this paper.
\vspace{-5pt}
\subsection{User Study}
We conducted a user study with simulated traffic scenes to collect multimodal data paired with trust labels. During the experiments, the participants would observe six videos presenting driving scenes on an AV, either being a car or a motorized scooter~\cite{hunter2023future, mehrotra2023trust}, from the driver's perspective. They were asked to monitor the AVs and periodically rate their trust level in the AV.
%We include two mobilities for more diverse traffic scenarios.  

Our video stimuli, totaling 14 including two tutorials, were adapted from the urban traffic environment detailed in~\cite{hunter2023future}. The videos showcased various autonomous driving styles, accounting for aggressiveness and the presence or absence of proactive audio prompts.
%, on either a car or sidewalk mobility, on a 24-inch screen. % Figure \ref{fig:attention} shows examples of the simulated scenes. 
% Sidewalk mobility refers to motorized scooters, and we included it for more diverse traffic scenarios~\cite{hunter2023future}.  %and click the arrow buttons to mark their intention to takeover. The up, down, left, and right keys stand for accelerate, slow down, turn left, and turn right, respectively. The participants were informed that the button clicking would not intervene the automated drive, but would be recorded. 
Each video was manually divided into multiple segments based on traffic events, such as turns and passing. At the end of each segment, the video would pause and the participants were prompted to rate their trust level of the AV. Specifically, they rated their agreement to the statement ``based on my current travel event, I trust my automated mobility'' on a Likert scale from 1 (Strongly Disagree) to 7 (Strongly Agree). Each participant was shown two tutorials and four random videos, during which their eye gaze data was captured with a Tobii Pro Nano eyetracker. We didn't use other physiological signals such as EEG and GSR, as their collection process is invasive and less stable~\cite{begum2013intelligent}, and thus is harder to deploy in applications.

A total of 48 subjects participated in the study. We excluded four participants for analysis, three due to a lack of attention and one due to data loss during the study. The remaining were 44 participants (mean age=28.6, SD=9.6) with 22 females. The entire experimental procedure took about one hour and participants were compensated with a \$20 gift card. The study was approved by the San Jose State University Institutional Review Board.
\vspace{-5pt}
\subsection{Data and Modalities}
\label{data-modalities}\vspace{-5pt}
Our dataset comprises 1,682 samples (segments) from 44 subjects and 12 different videos (tutorials excluded). The segments are 59.4 seconds long on average, with the shortest being 16 seconds and the longest 112 seconds long. Time-series features are all downsampled to 10Hz and aligned by unix timestamp. Each sample contains signals from the following modalities:
\vspace{-1pt}
\begin{enumerate}[wide, labelwidth=!, itemindent=!, labelindent=4pt]
    \item \textbf{Mobility signals} (5-dim, 30 Hz): include mobility state features obtained from the simulator: velocity, angular velocity, steering angle, engine sound volume, and on/off of real-time proactive voice. 
    \vspace{-12pt}
    \item \textbf{User gaze signals} (5-dim, 64 Hz): include the gaze point coordinates on the screen and the pupil diameters of each eye. We also include a categorical feature to indicate the type of object the gaze point falls on (e.g., pedestrian, road, etc.), calculated with the coordinates and per-frame semantic segmentation from the simulator.
    \vspace{-12pt}
    \item \textbf{Driving mode} (4-dim): Other than the time-series signals, we have metadata associated with each segment: the mobility (car or sidewalk), driving mode (aggressive or defensive, whether proactive prompts are provided). We also include the order of watching as it could potentially be a confounder, suggested in previous work~\cite{ekman2017creating}.
    \item \textbf{Trust label}: Following previous work, we converted the trust ratings to sufficient (1) or insufficient (0) trust~\cite{ayoub2023real,mehrotra2023does}. We set the trust threshold to 5 (5 is sufficient). The dataset is imbalanced: only 14\% of samples represent insufficient trust. 
\end{enumerate}
% \vspace{-3pt}

%The automated drive is SAE level 2, which provides both longitudinal and latitude control~\cite{SAE2018TaxonomyInternational}. 
 %v More details about the traffic events, study design, and the simulation environment can be found in the previous work. We will make the dataset publicly available for research purposes, upon the acceptance of this paper. 

\vspace{-10pt}
\section{Selective Windowing Attention Network}
\label{sec:model}\vspace{-5pt}
SWAN consists of three key modules: 1) the limited-range self-attention module to efficiently learn the local context; 2) the windowing module to convert timestep-wise representations to window embeddings, and embed the salient information within each window; 3) the weighting module that assigns window weights based on their features. Figure \ref{fig:AWAN} shows the full structure of SWAN. Below we review the multi-head dot-product attention, serving as the underpinning of our network. We then describe the three modules and the full network structure.
\vspace{-5pt}
\subsection{Multi-head Scaled Dot-Product Attention}\vspace{-5pt}
First proposed in the Transformer model \cite{vaswani2017attention}, Scaled Dot-Product Attention takes in three matrices: Query ($Q$), Value ($V$), Key ($K$), calculates weights matrix $W$ by the relevance of different parts in $Q$ and $K$, and reweighs parts in $V$ with those weights:
\vspace{-5pt}
\begin{equation}
W(Q, K) = \frac{Q\cdot K^T}{\sqrt{d_K}}
\tag{1}
\label{eq:weights}
\end{equation}
\vspace{-4pt}
\begin{equation}
\mathrm{Attention}(W, V) = \mathrm{softmax}(W)\cdot V 
\tag{2}
\label{eq:attention}
\end{equation}

% The underlying idea is to generate representations with focus on the parts in $V$ that are most relevant to $Q$. For example, the dot product $QK^T$ is high where $Q$ and $K$ align well. 
This calculation can be extended to multiple attention heads with their output concatenated, allowing parallel computation and more diverse representation. 
% $$\mathrm{MultiHead} (Q, V, K) = \mathrm{Concat}(\mathrm{head}_1, ..., \mathrm{head}_h)W^O$$
% , where $\mathrm{head}_i = \mathrm{Attention}(Q\cdot W_i^Q, K\cdot W_i^K, V\cdot W_i^V)$, and all $W$'s are trainable weights of linear projections. 
For simplicity, we only describe our modules with one attention head.
\vspace{-5pt}
\subsection{Modules}\vspace{-5pt}
\label{sec:modules}
\begin{figure}[t]
\centering
  \includegraphics[width=0.7\linewidth]{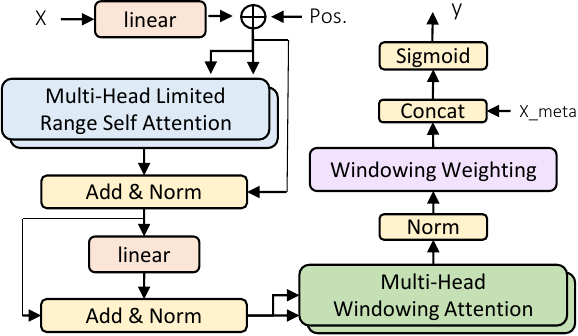}
  \vspace{-2pt}
  \caption{SWAN structure.}
  \label{fig:AWAN}
\end{figure}

\vspace{4pt}
\textbf{Limited Range Self-Attention.}
Similar to the encoder self-attention in the Transformer, we use a multi-head attention layer with $Q$, $V$, $K$ matrices being learned linear transformations of the input $X$ to embed contextual information. To better model the locality characteristic of our data, we apply a mask based on index distances to limit the attention of each signal step to its neighbors, with the self attention range $r_{self}$ being a hyper-parameter. Specifically, for an input X with size $L \times D$, where $L$ is the padded sequence length and D is the feature dimension, we calculate the self-attention mask $M_{self}$ of shape $L\times L$, where the element indexed by $i$, $j$ as:
\vspace{-2pt}
\begin{equation}
M_{self}[i][j] = (|i-j| <= r_{self})
\tag{3}
\end{equation}

Note that we refer to the attention window limit we give SWAM as a hyperparameter as ``window range'', to differentiate it from the exact ``window size'' we use in the traditional windowing approach. We apply the mask through element-wise multiplication on attention weights. Putting together, the self-attention layer performs the following transformation on X: 
\begin{equation}
\mathrm{SelfAtt}(X) = \mathrm{softmax}(\frac{(W_Q  X) (W_K X)^T}{\sqrt{D}} M_{self})(W_V X)
\tag{4}
\end{equation}

\textbf{Windowing Cross-Attention. }
The windowing attention layer then transforms the output from the self-attention layer from step sequences ($L \times D$) to window sequences ($L_{w} \times D$) through the use of dot-product attention with window masks and prompts. Given windowing range $r$ and step $s$, we create a windowing attention mask, $M_{w}$ ($L_{w} \times L$), to indicate the moving window ranges:
\vspace{-2pt}
\begin{equation}
M_{w}[i][j] = (s\times i <= j < s\times i + r)
\tag{5}
\end{equation}

We define window prompts $P$, as a matrix of shape $L_{w} \times D$ where $P[i]$ contains the prompt we use to query window $i$. Therefore, the transformation in \eqref{eq:weights} will generate an attention matrix of shape $L_{w} \times L$, indicating the attention weights assigned to each step by each window. Here, we use the max of step features within each window as the prompt for that window:
\begin{equation}
P[i'] = \underset{_i}{\mathrm{max}}(X[s\times i' : s\times i' + r])
\tag{6}
\end{equation}
\begin{equation}
\mathrm{WinAtt}(X) = \mathrm{softmax}(\frac{(W'_Q P) (W'_K X)^T}{\sqrt{D}}M_{w})(W'_V  X)
\tag{7}
\end{equation}

\textbf{Window Weighing. }
The windowing attention layer applies softmax in each window, making them sum up to 1. For the whole sequence, however, some windows may contain events that trigger trust change, and should be given more weights. We use a simple linear transformation matrix $W_w$ ($D\times 1$) to infer the window saliency from their embeddings, and calculate a full-sequence embedding as a weighted sum of window embeddings:
\begin{equation}
\mathrm{Output}(X) = (X \cdot W_{w}) \cdot X
\tag{8}
\end{equation}

\vspace{-8pt}
\subsection{Network Structure}\vspace{-5pt}
As Figure~\ref{fig:AWAN} shows, we first add positional encodings to the input X, as in~\cite{vaswani2017attention}, to help the attention layers learn the time information. Since the sequences are of variable lengths, we pad all sequences to a maximum length of 1500, and the actual lengths of each sequence are recorded and used as a padding mask to ensure those padded steps get zero weights. The input sequences are first passed through a Multihead Limited Range Self Attention layer with residual connections, followed by layer normalization and linear layers. They are then fed into a Multihead Windowing Attention layer, which outputs window embeddings. After another normalization layer, a Window Weighing layer aggregates the window embeddings into one embedding for the whole sequence.
% We then sequentially pass X through the three modules, with some layer-normalization, feed-forward layers and residual connection in between modules, following the structure in Transformer~\cite{vaswani2017attention}.
The learned embeddings are concatenated with sequence metadata, and are finally passed through a sigmoid layer to generate binary predictions. We train the model on Binary Cross Entropy loss with an Adam optimizer~\cite{kingma2014adam}.

\vspace{-2pt}
\section{Experiments}
\label{sec:result}\vspace{-5pt}
\subsection{Baselines}\vspace{-5pt}
We compare our model to three baselines: Random Forest (RF) with empirical windowing and feature extraction, CNN-LSTMs, and an encoder-only Transformer.\\
\textbf{Feature Extraction + RF.} 
Empirical feature extraction has been an effective and interpretable approach in many user state estimation tasks~\cite{zepf2020driver}. 
% In recent years, a lot of work proposed various handcrafted features paired with statistical or traditional machine learning models, such as Regressions, Support Vector Machines and Random Forests~\cite{lohani2019review}. 
We choose Random Forest (RF) as our classifier for the baseline since it's found to be effective and stable in relevant works~\cite{du2020predicting}. For feature extraction, we compiled a set of 50 features, covering a broad literature on our data modalities: for eye gaze signals, we extracted the statistical measurements (mean, standard deviation, min, max) of the gaze coordinates, pupil diameters, counts and lengths of fixation, dwelling and saccade, as well as saccade speed; for mobility state signals, we extract the statistical measurements of the velocity, angular velocity, steering angle and engine volume, and we also include one bit indicating whether proactive prompts occurred. We segment the input signals with a x-second moving window with y-second overlapping steps, and features are extracted on each window and aggregated through pooling. We report results with mean-pooling, since in our experiments it outperforms max-pooling.

\textbf{CNN-LSTM.} CNN-RNN has emerged as a popular model structure for modeling sequential data~\cite{canizo2019multi}. Within this framework, the convolutional layers act as feature extractors to identify meaningful local patterns, upon which the recurrent layers embed the sequential dynamics, enabling learning of both spatial and temporal characteristics. Yi et al. applied a CNN-LSTM network for multimodal feature fusion on the user trust prediction task, that achieved improved performance than SVM, CNN, and LSTM models~\cite{yi2023measurement}. Following common structures, our baseline model consists of two one-dimensional multi-channel convolutional layers, each followed by a ReLU activation layer and a max pooling layer. The output is then fed into an LSTM layer. Similar to SWAN, we concatenate metadata of each sequence to the last hidden state embeddings of the LSTM, and use another linear layer with a Sigmoid function to generate the predictions. Padding is also used in this model and ignored through masking during forward and backward propagation. 

\textbf{Transformer.} Although has not been applied to AV trust prediction, Transformer has achieved remarkable performance in many human state modeling tasks~\cite{vazquez2021using}. Since our task is binary classification, we applied a 6-layer encoder-only Transformer~\cite{vaswani2017attention}, connected to the same concatenation and classification layers as in SWAN. Since Transformer shares the same attention backbone as SWAN, using it as a baseline highlights the effect of our windowing layers.

\subsection{Experiment Setup}\vspace{-5pt}
We measure model performance through a subject-independent cross validation. We randomly divide subjects into 5 folds. For each run, we use samples from one fold for test, one for validation, and the other three for training. 
% In this way, we test the model capability to predict trust on new users, without observations of their behavioral responses. 
We use the validation set for model selection.
% (maximum depth and minimun decrease of impurity for RF; learning rate, kernel sizes and hidden size for CNN-LSTM; learningrate, number of attention heads and hidden size for AWAN) 
 For DNN approaches, we set all hidden dimensions to 10, the same as the input signal dimension. We train the model for 30 epochs, and use the model with the best validation performance for testing. To deal with the data imbalance, we upsample the minor class (insufficient trust) to match the size of the major class. We measure the models' performance with Unweighted Average Recall (UAR) since it gives equal importance to the classes and thus gives a clearer picture of performance on the imbalanced dataset~\cite{schuller09_interspeech}. We run all models with five random seeds, and report the mean and standard deviation of the performance across 25 runs (5 folds $\times$ 5 seeds).

\vspace{-5pt}
\section{Results and Analysis}

\begin{table}[t]
\centering
\small
\begin{tabular}{@{}lclc@{}}
\toprule
\multicolumn{1}{c}{} & window/kernel & \multicolumn{1}{c}{UAR} & \multicolumn{1}{c}{\#param} \\ \midrule
RF                   & 50                 & 0.615±0.08*             & /                           \\
CNN-LSTM             & {[}10, 3{]}        & 0.678±0.07*             & 1.6k                        \\
Transformer          & /                  & 0.685±0.09*             & 5.0k                        \\ \midrule
\textbf{SWAN}        & \textbf{30}        & \textbf{0.723±0.07}     & \textbf{1.3k}               \\
no SelfAtt           & 30                 & 0.715±0.07              & 1.3k                       \\
no WinAtt            & /                  & 0.693±0.06*             & 1.3k                        \\ \bottomrule
\end{tabular}
\caption{Trust Prediction Performance. We ran a paired t-test on the UAR across 5 folds and 5 seeds, a total of 25 runs, between SWAN and other baseline and ablation models. Asterix (*) indicates baseline worse than SWAN with significance p$<$0.05. }
\label{table-performance}
\end{table}

\begin{figure*}[t!]
\vspace{-10pt}
    \centering
    \includegraphics[width=.9\linewidth]{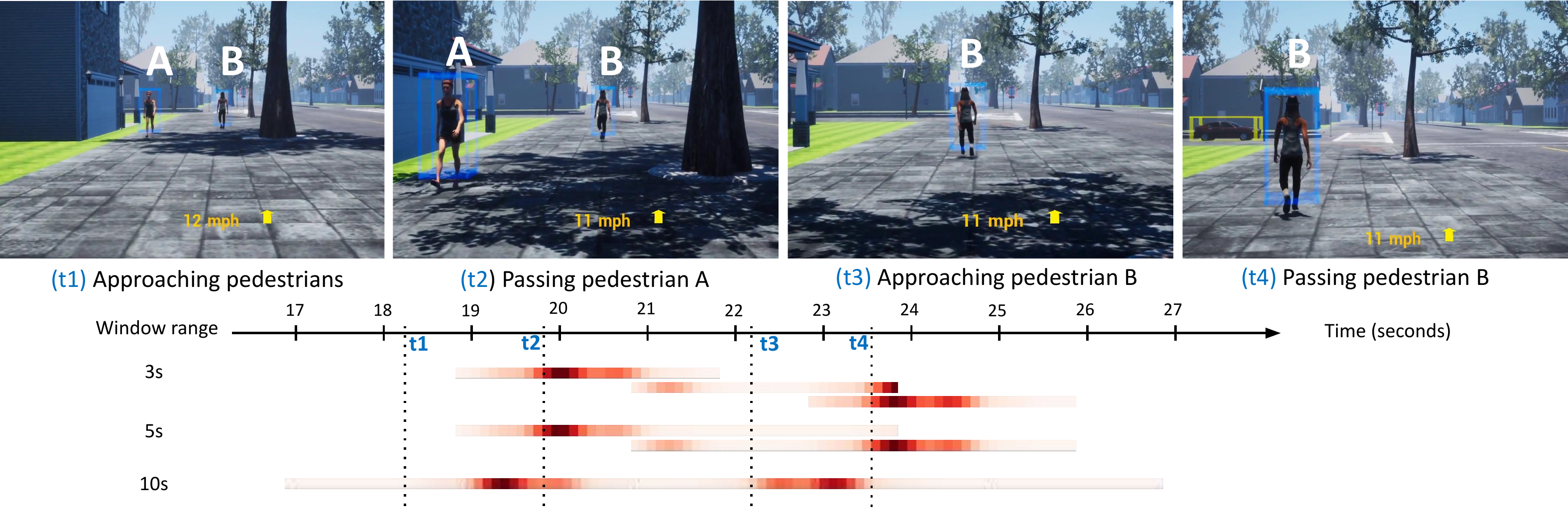}
    \vspace{-10pt}
    \caption{Four frames taken within 10 seconds from a sidewalk mobility video, and the model's attention weights visualizations on this interval.}
    \label{fig:attention}% \vspace{-3pt}
\end{figure*}

\begin{figure}[t]
        \centering
        \vspace{5pt}
        \includegraphics[width=0.6\linewidth]{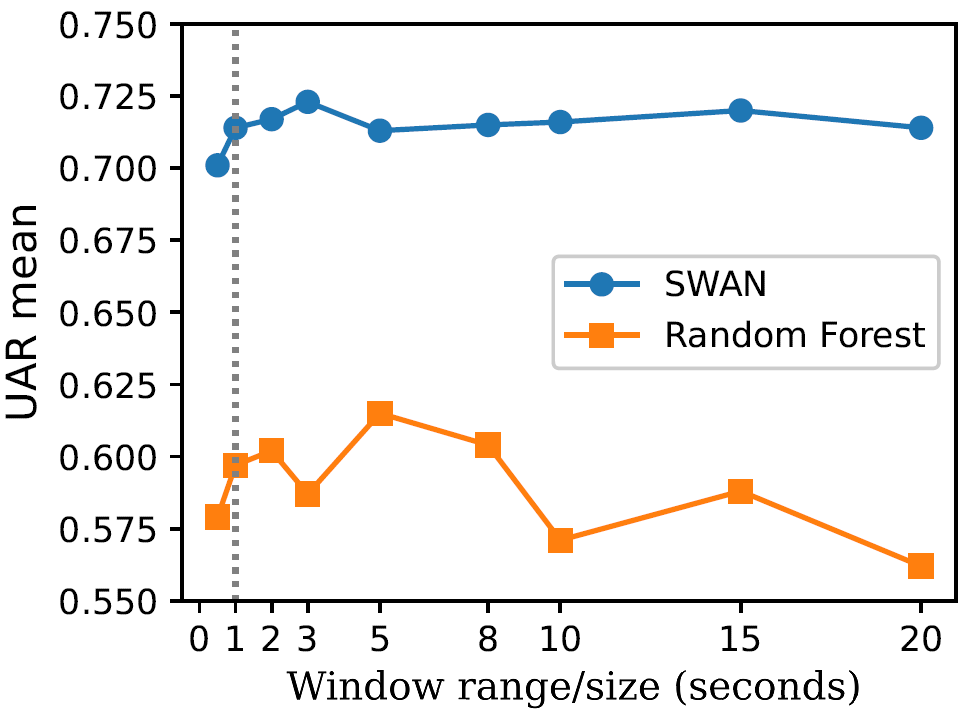}
        \vspace{-5pt}
        \caption{Performance across window ranges/sizes.} \vspace{-5pt} 
        \label{fig:windows}
\end{figure}

\vspace{-5pt}
\subsection{Trust Prediction Performance}\vspace{-5pt}

We compare SWAN with baseline models in Table \ref{table-performance}. Note that we performed hyperparameter selection on the validation set for each model, and the window ranges/kernel sizes listed are the best ones selected. SWAN achieves a 0.723 UAR on the binary trust prediction task, significantly (p$<$0.05) outperforms RF (0.615), CNN-LSTM (0.674) and Transformer (0.685) baselines. Note that the SWAN with the highest AUC was trained with a window range of 30 (3 seconds), while the best window size for Random Forest is 50. SWAN has a self-attention layer to encode context information before the windowing step, so a smaller optimal window range is reasonable.

We then perform an ablation study on SWAN modules. Table \ref{table-performance} shows that the Windowing Attention layer contributes significantly (p$<$0.05) to the performance. This indicates that, as a widely-adopted technique, windowing is beneficial for highlighting signal patterns within critical intervals, and that our end-to-end approach retains this advantage. The self-Attention layer also enhances the model's performance, albeit marginally.

\vspace{-5pt}
\subsection{Model Analysis}\vspace{-5pt}

\textbf{Attention weights.}
We first conduct a manual inspection on the weights assigned by the model within each windowing range. We use the model without self-attention to ensure a direct map from the highlighted positions to the corresponding intervals. For clearer visualization, we smooth the attention weights with a Gaussian kernel. Qualitatively, we find that the model tends to assign high weights to time periods when the AV has closer physical distances with other road users, specifically, when the AV is at risk of a collision if no action is taken by the automated mobility. As illustrated in Figure \ref{fig:attention}, the attention weights peak at t2 when the AV is getting close to pedestrian A, then recede once the AV initiates corrective action to maintain a safe distance. Interestingly, while the AV was in close proximity to A, it didn't receive heightened weights. This is logical because, despite the short distance from A, there is no expectation for the AV to take additional actions. The weights remain low until t4, when the AV has pedestrian B closely in front of it. This pattern of weight allocation suggests that the model's attention mechanism effectively identifies the connection between user trust and moments necessitating action.

\textbf{Window robustness.}
We then test how robust SWAN is across a wide span of Windowing Attention ranges. Aligning with previous work~\cite{konishi2022inferring}, we observe that the traditional window selection combined with the RF approach is sensitive to window size, as shown in Figure \ref{fig:windows}. Except for the sensitivity, the curve is not in a continuous convex shape, which implies that pinpointing the optimal window size usually requires a grid search, which is expensive. SWAN, on the other hand, has very robust performance with negligible variation (less than 0.01 in UAR) across windows ranged from 1 second to 20 seconds. The performance of SWAN drops when the window range is 0.5 second, which may be too short to contain sufficient information related to trust. Those results indicate that SWAN is insensitive to window range as a hyperparameter within a reasonable, wide enough range, and thus has the potential to save a lot of window selection efforts in human state detection applications. 

To further understand SWAN's robustness across window sizes, we visualize the attention weights across models trained with Windowing Attention ranges of 3, 5, and 10 seconds, using the segment presented in Figure \ref{fig:attention}.  The attention weights generally align well across window sizes, with small temporal discrepancies for the 10s window model. The observed pattern further signifies the adaptability and robustness of the windowing attention layer.
% that SWAN is able to identify the pivotal intervals associated with trust shifts. 
% This stability across intervals verifies that SWAN is able to operate like human beings, paying attention to critical travel events.
% \vspace{-5pt}
\subsection{Discussion}
\vspace{-3pt}
SWAN has a few advantages. First, compared to traditional windowing methods, SWAN allows flexible weight assignment and thus more accurately captures the critical intervals of trust change.
% Fixed-size windowing can lead to information loss in many ways: an overly small window could cause pattern truncation (e.g., eye gaze saccades), while one too large could dilute signals (e.g. the mean of a signal across a long time span may appear stable, even if some peaks occur in the sample). 
Besides, the attention backbone ensures good interpretability. % We can probe where the model assigns higher weights and assess if that aligns with human sense.

SWAN is also lightweight and effective in learning from small data. Compared to convolutional layers where kernel size is a factor of the number of parameters, 
% AWAN has fewer parameters compared to CNN: given a window size $r$ and input and output feature dimensions both being $D$, a convolution layer would have $r \times D \times D$ parameters, while one attention layer in AWAN has $C \times D \times D$, where $C$ is a constant number of the linear transformations in the attention layer, in our case being four ($W_Q, W_K, W_V$, and the output linear layer). Since AWAN's parameter size is not dependant on the window length, 
SWAN's attention layer only has a constant number of linear transformations (in our case four: $W_Q, W_K, W_V$, and the output linear layer). This makes SWAN more scalable, especially for signals with long patterns or high sampling rates. Compared to the vanilla dot-product attention where pair-wise similarity is computed across all time steps, SWAN essentially reduces computation complexity through generous masking: the windowing attention layer focuses on a small region at a time, masking all other time steps. Although this doesn't reduce the number of parameters in the model, it greatly reduces the effective input dimensions involved in the back propagation, allowing localized patterns to emerge. 
% Existing work has reached prediction in the region of 80-81\%,~\cite{yu2021measurement, ayoub2022real}, compared to xx.x\% from this work. We emphasize that our data does not include wearable measurements such as physiological and haptic detection, which may be invasive in real-world applications. 
It's possible that, given sufficient training data, models with multiple stacked pairwise attention layers such as the Transformer may still outperform SWAN with their enhanced modeling capacity. However, given the high cost of acquiring human state labels and the typically limited size of such datasets, SWAN stands out as a valuable tool for many tasks.

\vspace{-8pt}
\section{Conclusion}\vspace{-8pt}
This research addresses the challenge of modeling human internal state changes derived from prolonged multimodal time-series signals, using trust prediction in autonomous mobilities as the case study. We introduce SWAN, an attention-based neural network utilizing range masks for adaptive windowing. Our experimental results show SWAN's proficiency in identifying crucial intervals of human state changes, leading to a significant enhancement in trust prediction performance. It also displays strong robustness across a diverse spectrum of window ranges, alleviating the tedious process of searching for the optimal window size. We compare SWAN to RF, CNN-LSTM, and Transformer baselines, highlighting its interpretability and light weight. Finally, we discuss the task and data types that best align with its modeling capabilities.

\vspace{-5pt}
\section{Acknowledgement}\vspace{-5pt}
The authors would like to sincerely thank Prof. Anil Kumar, Elise Ulwelling, Jessie Snyder, and Mayank Srinivas from San Jose State University for their help in data collection.

\footnotesize
\bibliographystyle{IEEEbib}
\bibliography{refs}

\end{document}